\documentclass[12pt,preprint]{aastex}

\newcommand{\msun}{{~{\rm M}_\odot}}

\def\arcsec{\hbox{$^{\prime\prime}$~}}

\shorttitle{Star Formation Rates in Cooling Flow Clusters}
\shortauthors{Hicks \& Mushotzky}

\begin{document}

\title{Star Formation Rates in Cooling Flow Clusters: A UV Pilot Study with Archival XMM-Newton Optical Monitor Data}
\author{A.K. Hicks\altaffilmark{1} and R. Mushotzky\altaffilmark{2}}
\altaffiltext{1}{Center for Astrophysics and Space Astronomy, University of Colorado at Boulder, Campus Box 389, Boulder, CO 80309; ahicks@alum.mit.edu}
\altaffiltext{2}{Goddard Space Flight Center, Code 662, Greenbelt, MD, 20771; richard@milkyway.gsfc.nasa.gov}

\begin{abstract}

We have analyzed XMM-Newton Optical Monitor (OM) UV (180-400 nm) data for a sample of 33 galaxies.  30 are cluster member galaxies, and nine of these are central cluster galaxies (CCGs) in cooling flow clusters having mass deposition rates which span a range of $8-525 \msun~\rm{yr}^{-1}$.  By comparing the ratio of UV to  2MASS J band fluxes, we find a significant UV excess in many, but not all, cooling flow CCGs, a finding consistent with the outcome of previous studies based on optical imaging data \citep{mcnamara1,cardiel,crawford}.  This UV excess is a direct indication of the presence of young massive stars, and therefore recent star formation, in these galaxies.  Using the Starburst99 spectral energy distribution (SED) model of continuous star formation over a 900 Myr period, we derive star formation rates of $0.2-219 \msun~\rm{yr}^{-1}$ for the cooling flow sample.  For 2/3 of this sample it is possible to equate Chandra/XMM cooling flow mass deposition rates with UV inferred star formation rates, for a combination of starburst lifetime and IMF slope.  This is a pilot study of the well populated XMM UV cluster archive and a more extensive follow up study is currently underway. 

\end{abstract}

\keywords{cooling flows---galaxies:clusters:general---galaxies:elliptical and lenticular, cD---galaxies:stellar content---stars:formation---ultraviolet:stars}

\section{Introduction \label{s:intro}}

The 'mystery' of cooling flows \citep{fabian1,sarazin,fabian2} has baffled astronomers for over 20 years. Though the X-ray temperature often drops by a factor of three within the central 100 kpc \citep{allen}, there is seldom enough mass at temperatures cooler than $\sim2$ keV to be consistent with constant pressure cooling flow models.  Recent XMM and Chandra results show that the distribution of temperature vs. emission measure also does not match the prediction of these models \citep{peterson,voigt}. While the XMM and Chandra results have reduced the implied cooling rate in most clusters there is still an apparent discrepancy of an order or magnitude or more between the X-ray cooling rates and the observed star formation rates (SFR) in many clusters. Contrary to conventional wisdom that there is no star formation at all in the central cluster galaxies, a number of contemporary studies have found significant star formation \citep{melnick,smith,hansen,mittaz,mcnamara2,odea,mcnamara3}. However we are still far from a complete reconciliation between X-ray mass deposition rates and alternatively derived star formation rates in the majority of cooling flow clusters. 

While many cluster cooling flows have been well observed in the radio and with emission line studies \citep[and others]{heckman} there have been very few published studies of the UV properties of cooling flow clusters.  In this study we use the readily available archival XMM Optical Monitor UV observations to derive the UV luminosities of a moderate sample of clusters. Since the dominant contribution to UV flux comes from short-lived main sequence stars, and the total number of these stars is directly proportional to the star formation rate \citep{kennicutt} the UV is one of the prime routes to understanding star formation. We attempt to address and quantify the connection between cooling flows and star formation in our pilot sample. Unless otherwise noted, this paper assumes a cosmology of $\rm{H}_0=70~\rm{km}~\rm{s}^{-1}~\rm{Mpc}^{-1}$, $\Omega_{\rm{M}}=0.3$, and $\Omega_{\Lambda}=0.7$

\section{The XMM-Newton Optical Monitor}

The XMM Optical/UV Monitor Telescope is a modified 30cm Ritchey-Chretien telescope mounted alongside the X-ray mirror modules \citep{mason} which allows simultaneous X-ray and UV/optical observations of XMM targets within the central 17 square arcminute region of the X-ray field of view.  The $256\times256$ pixels have an angular resolution of $\sim 1 \arcsec$.  The sensitivity limit of the Optical Monitor (OM) is 24 mag. This study primarily utilizes the UVW1 filter (220-400 nm), and the UVW2 filter (180-260 nm), and the UVM2 filter (200-280 nm) for calibration purposes. Previous XMM OM studies of clusters have been limited to the detection of UV mission in Abell 1795 \citep{mittaz}. 

\section{Sample and Observations}

\placetable{table1}

Our pilot sample consists of 33 galaxies and eleven clusters.  Three of them (M49, M60, and M86) were selected because they possess IUE photometric information \citep{burstein} and were used to compare OM to IUE fluxes.  21 galaxies in the IC 1860 and Coma fields were used to calibrate the relationship between UV and IR luminosities in passively evolving early type and cD galaxies.  Two UVW1 observations of Coma were included, one centered on the cD galaxies, and one which is off-center and contains numerous early-type galaxy detections within the viewing field.  

The remaining nine clusters in our sample were chosen because of significant X-ray determined mass deposition rates, general overlap with the samples of \citet{mcnamara1} and \citet{cardiel}, moderate to low redshift, inclusion in the 2MASS galaxy sample, and having archival XMM UVW1 data.  The majority of clusters in our sample are low redshift objects ($z<0.1$), but we have also included three moderate redshift clusters ($z\sim0.25$) which are known to possess very strong cooling flows.

\begin{figure}
\plotone{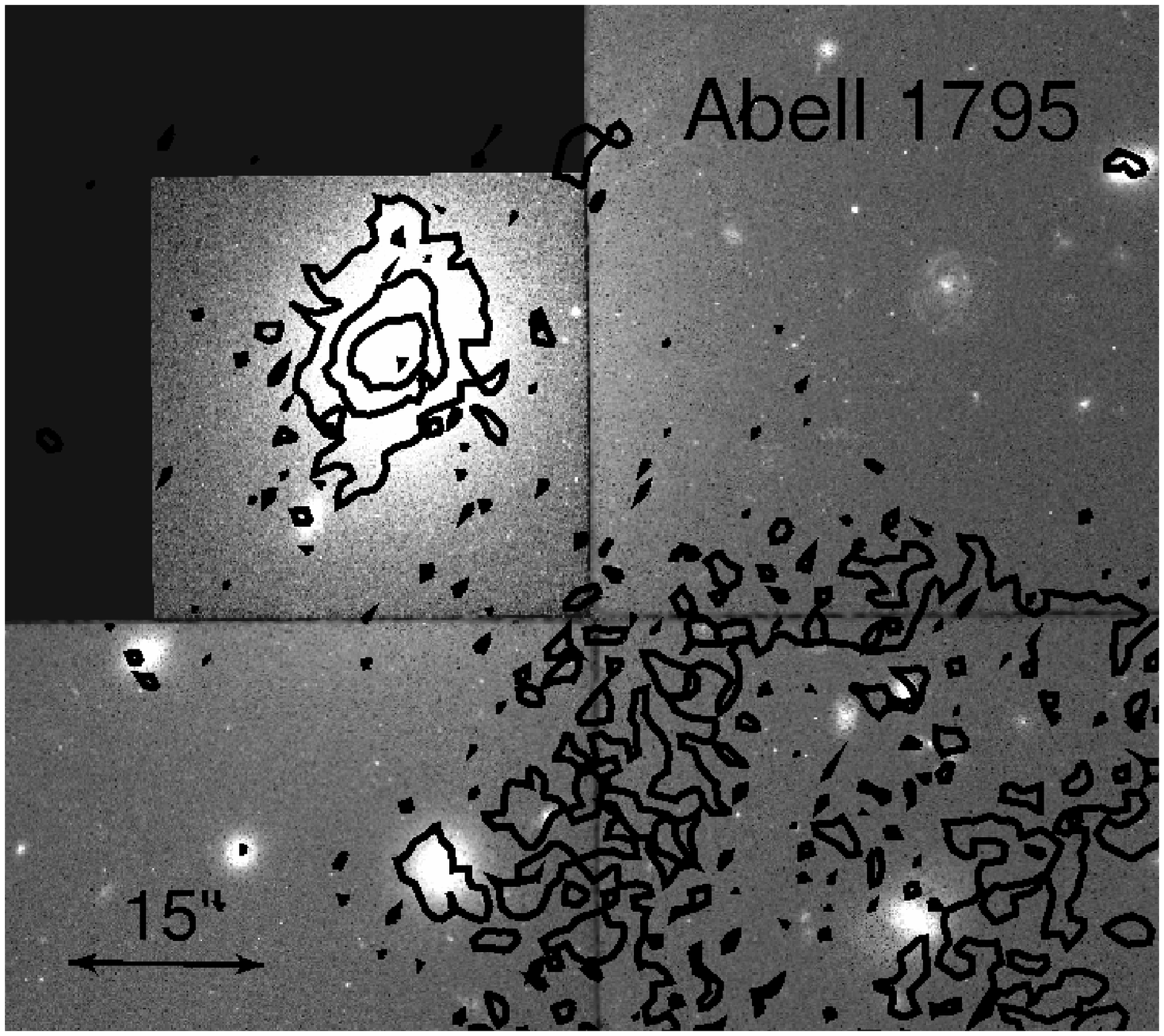}
\vspace{-30mm}
\caption{HST WFPC2 1780 second exposure image of Abell 1795.  This image has been overlaid with UV contours from an XMM-Newton Optical Monitor UVW1 filter (220-400 nm) observation.  Logarithmic contours in black were produced with a 2499 second exposure image, and have count values of 14, 25, 44.75, and 80.  The large splotch of UV contours to the lower right of the cD galaxy is due to scattered light.\label{fig1}}
\end{figure}

Figure \ref{fig1} shows one of the cooling flow clusters in our sample, with XMM-Newton UVW1 contours overlaid.  Table \ref{table1} lists the objects in our current sample, their redshifts, relevant filters, and OM exposure times.  

\placefigure{fig1}

\section{Surface Brightness\label{sb}}

We fitted the flux in concentric circles around the emission peak of each galaxy to a radial surface brightness profile of the form:

\begin{equation}
I(r) = I_0 \left( 1 + {r^2 \over r_0^2} \right)^{1/2 - 3\beta} + I_B
\label{sb_eq}
\end{equation} 

\noindent Where $I_{0}$ is the normalization, $r_{0}$ is the core
radius, and $I_B$ is a constant representing the intensity
contribution of the background.

\begin{figure}
\includegraphics[angle=90,width=6.5in]{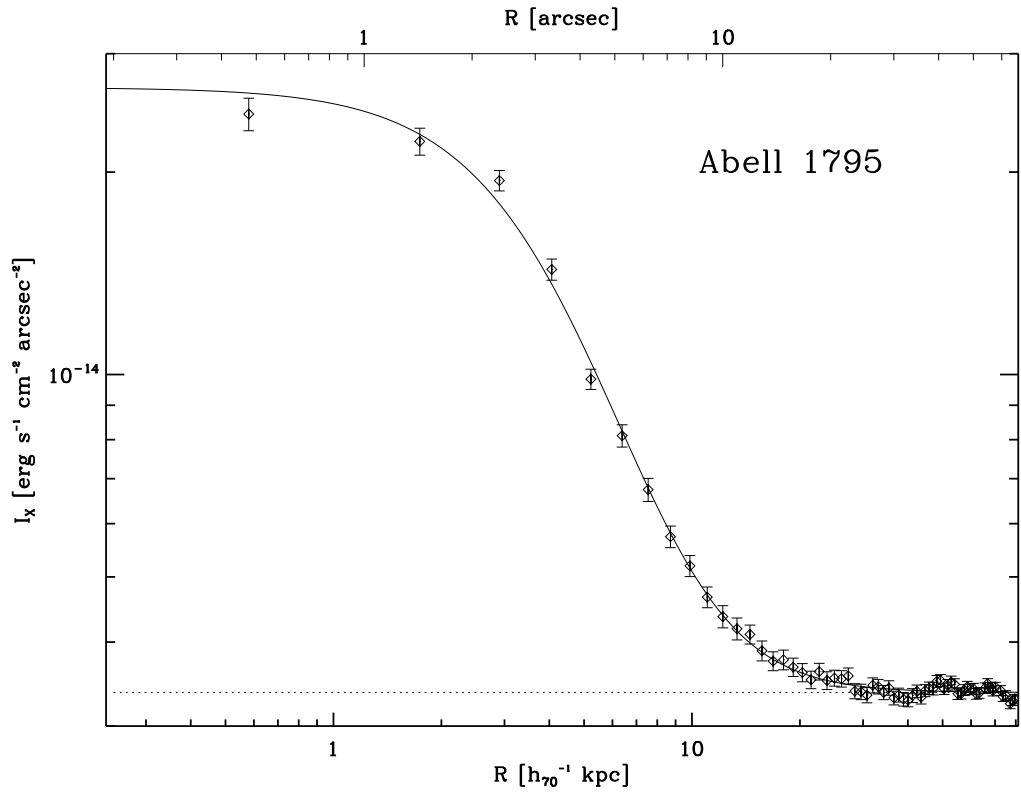}
\caption{{\bf Surface Brightness Profile.} Radial surface
brightness profile of Abell 1795 made from UVW1 (220-400 nm) data accumulated in $\sim1\arcsec$ annular bins.  A solid line traces the best $\beta$ model fit.  The horizontal dotted line
represents the best fitting background value.  The best fitting $\beta$ model has a core radius of 4.09 $\arcsec$ and beta value of 0.703 \label{fig2}}
\end{figure}

The surface brightness profiles are well fit by these models, cf. Figure \ref{fig2}.  The fit parameters are in Table \ref{table2}.

\placefigure{fig2}

\placetable{table2}

\section{Photometry}

We retrieved pipeline-processed UV data from the XMM-Newton Science Archive and combined observations with the same pointing. Flux determinations for our calibration sample used a circular aperture of 7\arcsec radius, a good approximation of the IUE aperture \citep{burstein}, and the size of the fixed 2MASS photometric aperture.  Photon counts in the aperture were corrected for background using a 35\arcsec radius background region taken from an outlying area of the chip in the same observation.  Final fluxes were determined by employing the OM counts-to-flux conversions, and errors were assessed via $\sqrt{\rm{N}}$ photon statistics.  

The results from UVW2 and UVM2 observations of M49, M60, and M86 were then compared with the IUE determined flux value of each object \citep{burstein} for the bandpass of the appropriate filter.  Our flux measurements are consistent with theirs to within $<35\%$, errors which are in keeping with those of \citet{vass} and \citet{koratkar}.  Two of the three galaxies have lower OM fluxes than IUE fluxes, indicating that the Optical Monitor does not have a red leak.  

We then determined UVW1 fluxes for the remainder of our sample (see Table \ref{table1}), for a 7\arcsec radius region, in the method described above. The UVW1 data were corrected for average galactic extinctions \citep{cardelli}.  J band extinction values were retrieved from the NASA/IPAC Extragalactic Database (NED).

\section{Calibration\label{cal}}

The low redshift ($z\sim0.023$) calibration sample has 23 galaxies in IC 1860 and Coma with 2MASS J band photometric measurements.  Two of these galaxies exhibited high UV fluxes with respect to the rest of the calibration sample, and were described as star forming in a FOCA study \citet{donas} and thus are not appropriate for inclusion in our UV/IR luminosity calibration of passively evolving early type galaxies, and were removed from the remainder of this study.

A least squares fit between the UVW1 and J band luminosities for the remaining 21 early type galaxies is well described by 

\begin{equation}
{\rm{L}}_{\rm{UV}} = C_1 \times {\rm{L}}_{\rm{IR}}^{~C_2}
\end{equation}

\begin{figure}
\includegraphics[angle=90,width=6.5in]{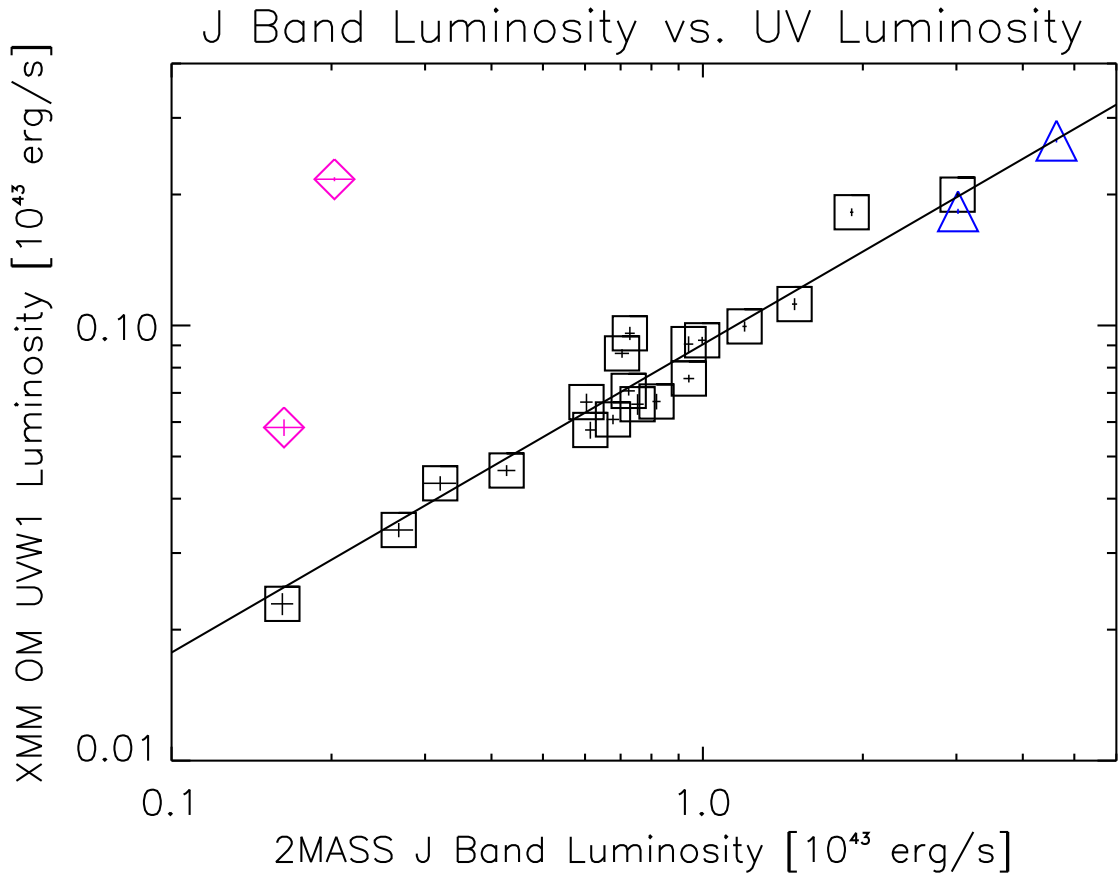}
\caption{UV vs. IR luminosity for non-cooling flow cluster galaxies in our sample.  The line indicates the best fitting relationship between UV and IR luminosity for passively evolving and non-cooling flow cD galaxies, represented by squares and triangles, respectively.  Diamonds indicate the two star forming galaxies \citep{donas} which were excluded from the fit.  A relationship between UV and IR luminosity is expected here, because passively evolving cluster galaxies have similar spectral energy distributions (SEDs).  The similar redshifts of IC1860 and Coma allow us to fit this relationship without making distance corrections to the data.\label{fig3}}
\end{figure}

Where ${\rm{L}}_{\rm{UV}}$ is the UVW1 luminosity and ${\rm{L}}_{\rm{IR}}$ is the 2MASS J band luminosity, inside 7\arcsec radius apertures with $C_1$ = $0.090\pm{0.007}$ and $C_2$ = $0.708\pm{0.007}$ respectively (Figure \ref{fig3}).  We use this correlation to predict the UV light from non-starforming, passive galaxies assuming that the UV light from such objects is produced by old evolved stars coming from the same population that produces the IR light.

\placefigure{fig3}

\section{Photometric Results\label{results}}

\begin{figure}
\includegraphics[angle=90,width=6.5in]{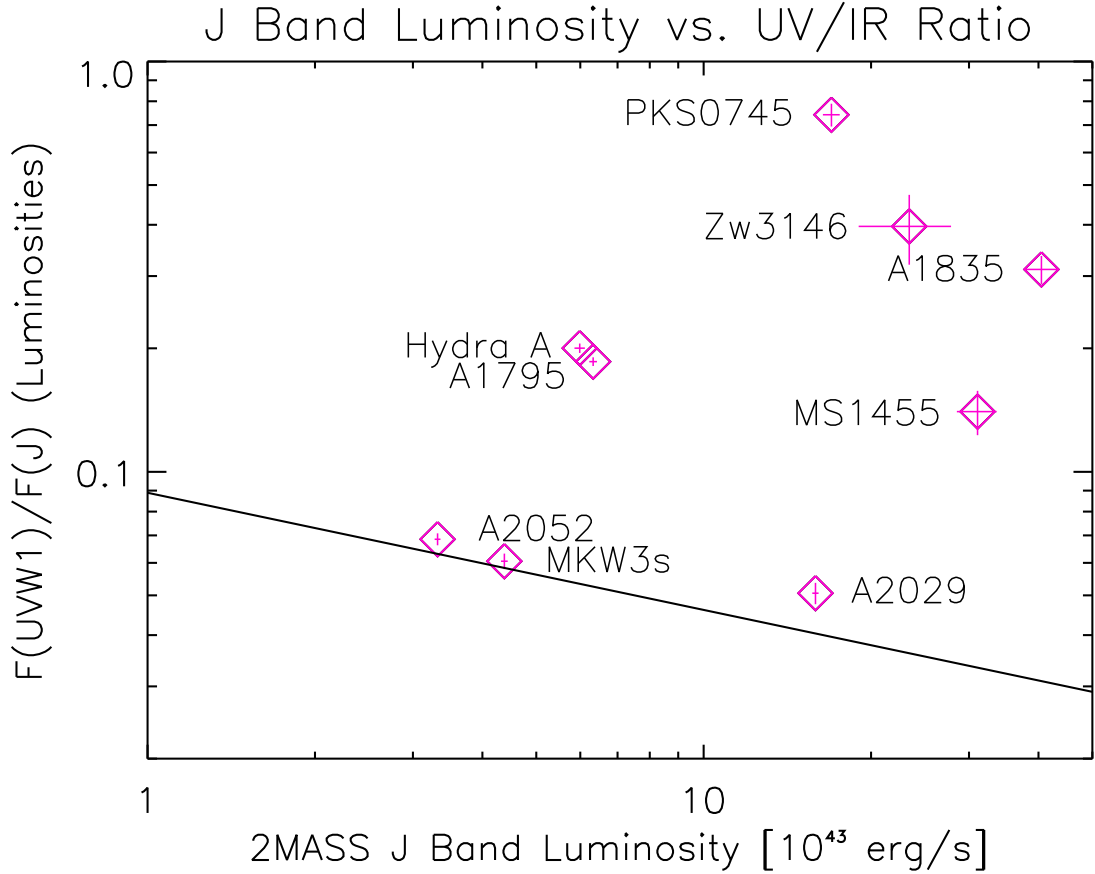}
\caption{UV/IR luminosity ratio of selected galaxies vs. their 2MASS J band luminosities.  All luminosities were calculated in $7\arcsec$ radius regions, and corrected for galactic extinction.  Diamonds indicate cooling flow cluster cDs, all of which are at low redshift ($z<0.1$) except for the three points farthest to the right, which have redshifts of $\sim0.25$.  The line designates the best fitting relationship between UV and IR luminosity for passively evolving and non-cooling flow cD galaxies (Section~\ref{cal}).  The high UV/IR ratio of the cooling flow clusters is a strong indication of the existence of hot young stars and thus a direct indication of copious recent star formation.\label{fig4}}
\end{figure}

The UV luminosity excesses of the cD galaxies in our nine cooling flow clusters were calculated by subtracting the expected UV light due to old stars (based on the fit obtained in Section \ref{cal}) from the measured UV luminosity of each cluster.  Figure \ref{fig4} shows the J band luminosity of each galaxy in our sample plotted against its UV/IR luminosity ratio.  We have not attempted to estimate and correct for internal dust absorption, and thus these excesses provide a lower limit on the UV emission in these clusters.  

\placefigure{fig4}

Because we are analyzing data from a single filter, it is difficult to break the degeneracy between age and mass of the young stellar populations. We folded through the UVW1 filter Starburst99 \citep{leitherer} redshifted models corresponding to continuous star formation over a a variety of timescales with solar abundance ($\rm{Z}=0.020$) and IMF powerlaw indices of both 2.35 and 3.3 to establish the relation between OM UV counting rates and star formation rate. The measured UV luminosity excesses of the clusters were then used to estimate per-year star formation rates (Table~\ref{table3}) that were then compared to X-ray derived mass deposition rates from \citet{voigt}, \citet{peterson}, and R. Mushotzky (2005, private communication).  Also in Table \ref{table3} are estimates of the duration of continuous star formation, and best fitting IMF powerlaw index.  These values were derived by plotting the relationship between star formation duration and rate for each given UV excess and IMF slope, then finding the set of parameters (slope and duration) at which X-ray determined mass deposition rates intercepted those curves.
The 900 Myr models should be closest to the average lifetime of a cooling flow \citep{allen}.
 
\placetable{table3}

\section{Summary and Discussion \label{discussion}}

We find a significant UV luminosity excess in the central galaxies of most cooling flow clusters.  This implies that star formation has recently been or is currently occurring most likely as a direct result of radiative cooling of the intracluster medium (ICM).  These conclusions are qualitatively consistent with the findings of previous optical imaging studies of star formation in cooling flow clusters \citep{mcnamara1,cardiel,crawford}.  

In the absence of multiwavelength spectral information, it is difficult to constrain a particular model of the star formation history of these galaxies.  However we note that
 for 2/3 of our sample it is possible to find some combination of the parameters of IMF slope and starburst lifetime which allow us to equate X-ray determined mass deposition rates with UV inferred star formation rates.

In none of the OM, Chandra or HST images is there evidence of a significant excess central flux, therefore we expect the AGN UV contribution to our signal is small.  In addition, we have investigated the amount of UV flux which might come from the cooling flow itself, and have found it to be much smaller than measured UV excesses.  Since the calibration sample consists entirely of low redshift early type galaxies, there is a possible systematic error in our calculations of the SFRs of the three moderate redshift ($z\sim0.25$) clusters in our sample, due to evolution in the UV luminosity of the old stellar population \citep{brown} and/or uncertainties in the redshift evolution of the star forming component.  In future work we will attempt to calibrate a higher redshift passive E galaxy sample.

\begin{figure}
\includegraphics[angle=90,width=6.5in]{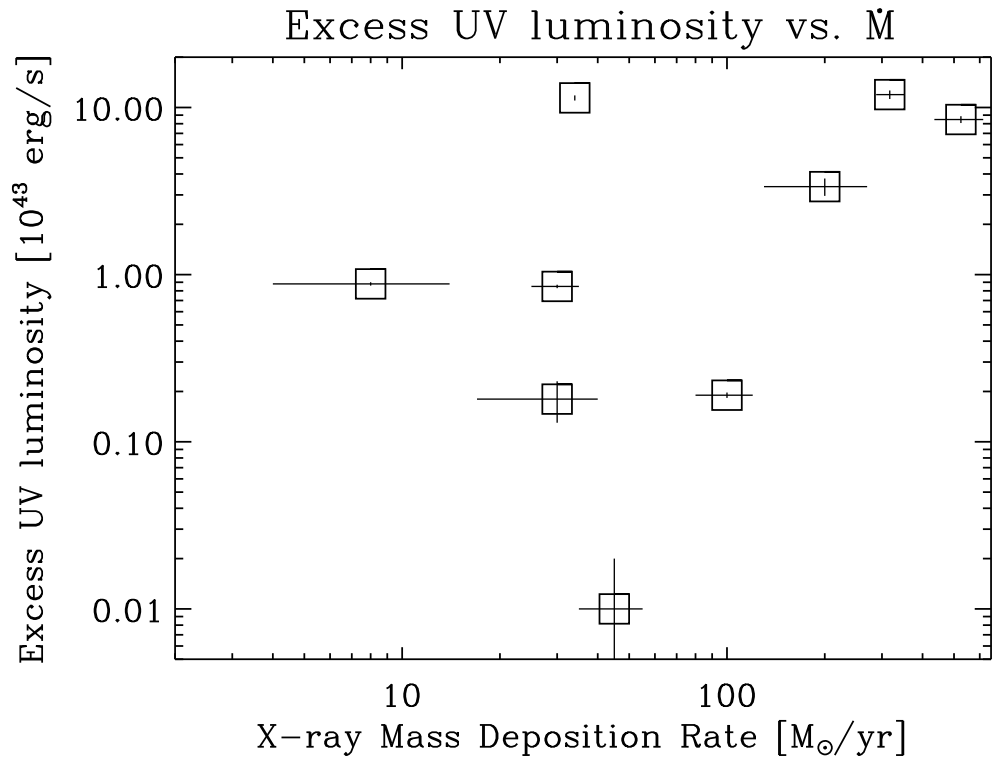}
\caption{Excess UV luminosity vs. $\dot{\rm{M}}$.  Luminosity excesses were determined by subtracting the best fit ratio of UV light due to old stars (Figure \ref{fig2}) from the measured UV luminosity of each cluster. X-ray mass deposition rates were taken from recent papers \citep{peterson,voigt}. \label{fig5}}
\end{figure}

While there is significant UV flux in 2/3 of our cooling flow sample, 1/3 of the objects with significant cooling rates from modern data do not have detectable UV excesses, and the ratio of UV light to X-ray cooling rate (Figure \ref{fig5}) shows a wide range. The origins of this scatter are unclear and a larger UV sample, with direct comparison of X-ray and radio imaging and IR data, is necessary to get a clearer picture of the origin of the UV light. The discovery of copious UV emission from several cooling flow clusters confirms the previous work and shows that a significant fraction of the cooling gas in clusters can indeed form stars. 

\placefigure{fig5}

Our star formation estimates are very basic, being based solely on a simple conversion from UV light to SFR \citep{kennicutt}. We have bypassed issues of the age of the star formation, the effects of reddening, the slope of the IMF and metallicity effects. The fact that we can, roughly, obtain similar values of X-ray inferred cooling rate and UV inferred star formation rate indicates that, given the above degeneracies, more and better data (such as could be obtained by Galex or HST spectroscopy) is necessary to determine precise star formation rates.

This paper is a pilot study and we are currently undertaking an expansion of this project toward higher redshift, utilizing more of the OM UV filters and including the many additional clusters that remain in the archive.  Through this expansion we hope to achieve a more thorough characterization of recent star formation in cooling flow clusters of galaxies.

\acknowledgements 

This work was made possible by a NASA Graduate Student Research Program (GSRP) fellowship, NGT5-140.

This research has made use of the NASA/IPAC Extragalactic Database (NED) which is operated by the Jet Propulsion Laboratory, California Institute of Technology, under contract with the National Aeronautics and Space Administration.













\clearpage

\clearpage
\begin{deluxetable}{ccccc}
\tablecolumns{5}
\tablewidth{0pt}
\tablecaption{XMM-Newton Optical Monitor Observations\label{table1}}
\tablehead{
\multicolumn{2}{c}{Object} & 
\colhead{{\rm{z}}} & 
\colhead{Filter} &
\colhead{Exposure [s]}}
\startdata
\multicolumn{2}{l}{IC 1860} & 0.0229  &UVW1 &5000\\
\multicolumn{2}{l}{PKS 0745-191} &  0.1028  &UVW1 & 3999\\
\multicolumn{2}{l}{Hydra A} &  0.0538 &UVW1  & 3000\\
\multicolumn{2}{l}{Zw3146} &  0.2906 &UVW1  & 4000\\
\multicolumn{2}{l}{NGC 4406 (M86)} & -0.0008  &UVW1& 3600 \\
\multicolumn{2}{l}{} &  & UVM2  & 3500 \\
\multicolumn{2}{l}{NGC 4472 (M49)} & 0.0033 &UVW1& 2000\\
\multicolumn{2}{l}{} &  & UVM2  & 1000 \\
\multicolumn{2}{l}{NGC 4649 (M60)} & 0.0037 &UVW1 & 2499\\
\multicolumn{2}{l}{} &  & UVW2  & 2900 \\
\multicolumn{2}{l}{Coma (center)} & 0.0231 &UVW1 & 3900\\
\multicolumn{2}{l}{Coma (outskirts)} & &UVW1 &1498\\
\multicolumn{2}{l}{Abell 1795} &  0.0633   &UVW1& 2499\\
\multicolumn{2}{l}{Abell 1835} &  0.2520  &UVW1 & 2000\\
\multicolumn{2}{l}{MS 1455.0+2232} &  0.2578 &UVW1  & 2000\\
 \multicolumn{2}{l}{Abell 2029} & 0.0779 &UVW1  & 1200\\
\multicolumn{2}{l}{Abell 2052} &  0.0345&UVW1  & 2000\\
\multicolumn{2}{l}{MKW3s} &  0.0453  &UVW1 & 2002\\
\enddata 
\end{deluxetable}

\clearpage
\begin{deluxetable}{lccccc}
\tablecolumns{6}
\tablewidth{0pc}
\tablecaption{Beta-Model Fits\label{table2}}
\tablehead{ 
\colhead{Object} & \colhead{$R_{\rm core}$} [\arcsec] & \colhead{$\beta$}  & \colhead{$I_{\rm 0}$\tablenotemark{a}} & \colhead{$I_{\rm B}$\tablenotemark{a}}  & \colhead{Reduced $\chi^2$} } 
\startdata 
{PKS 0745-191} & $2.83^{\pm{0.04}}$ & $0.766^{\pm{0.011}}$ &
$625.3^{\pm{12.4}}$ & $481.0^{\pm{1.6}}$ & 0.691  \\ 

{Hydra A} & $3.449^{\pm{0.001}}$ & $0.8387^{\pm{0.0002}}$ &
$4829.1^{\pm{2.4}}$ & $356.59^{\pm{0.06}}$ & 4.200  \\ 

{Zw3146} &  $6.21^{\pm{0.01}}$ & $2.054^{\pm{0.007}}$ &
$1119.2^{\pm{4.3}}$ & $413.2^{\pm{0.4}}$ & 0.947 \\ 

{Abell 1795} & $4.09^{\pm{0.01}}$ & $0.703^{\pm{0.002}}$ &
$2334.8^{\pm{13.4}}$ & $336.6^{\pm{0.3}}$ & 0.946 \\ 

{Abell 1835} &  $3.65^{\pm{0.41}}$ & $1.253^{\pm{0.165}}$ &
$2880.7^{\pm{153.0}}$ & $349.6^{\pm{1.8}}$ & 0.514 \\ 

{MS1455.0+2232} &   $4.99^{\pm{0.03}}$ & $1.006^{\pm{0.007}}$ &
$446.9^{\pm{3.5}}$ & $337.8^{\pm{0.4}}$ & 0.657\\ 

{Abell 2029} &  $3.68^{\pm{0.02}}$ & $0.499^{\pm{0.001}}$ &
$613.2^{\pm{3.5}}$ & $449.7^{\pm{0.6}}$ & 1.101\\ 

{Abell 2052} &  $1.68^{\pm{0.28}}$ & $0.402^{\pm{0.022}}$ &
$1593.1^{\pm{149.6}}$ & $426.2^{\pm{7.4}}$ & 1.912\\ 

{MKW3s} &  $1.28^{\pm{0.33}}$ & $0.352^{\pm{0.022}}$ &
$1018.0^{\pm{138.9}}$ & $398.6^{\pm{7.1}}$ & 1.110\\
\enddata
\tablenotetext{a}{Surface brightness $I$ in units of $10^{-17}$ ergs sec${}^{-1}$ cm${}^{-2}$ arcsec${}^{-2}$}
\end{deluxetable}

\clearpage
\begin{deluxetable}{ccccccccc}
\tablecolumns{9}
\tablewidth{0pt}
\tablecaption{\label{table3}}
\tablehead{
\multicolumn{2}{c}{Cluster} & 
\colhead{{\rm{z}}} & 
\colhead{UV Excess} & 
\colhead{$\dot{M}_{2.35}$\hspace{0.1cm}\tablenotemark{a}} &
\colhead{$\dot{M}_{3.3}$\hspace{0.1cm}\tablenotemark{a}} &
\colhead{$\dot{M}$ (X-ray)} &
\colhead{$\tau_{\rm{sf}}$} &
\colhead{$\alpha$} \\
\multicolumn{2}{c}{} & 
\colhead{}  & 
\colhead{[$10^{43}$ erg/s]} & 
\colhead{[$M_{\sun}$/yr]}& 
\colhead{[$M_{\sun}$/yr]}& 
\colhead{[$M_{\sun}$/yr]} &
\colhead{[Myr]} &
\colhead{}}

\startdata
\multicolumn{2}{l}{PKS 0745-191} &  0.1028  & $11.94^{\pm{0.7}}$ & $129^{\pm{7}}$ & $237^{\pm{13}}$ & $317^{+35}_{-29}$\hspace{0.1cm}\tablenotemark{b} & 200 & 3.3 \\
\multicolumn{2}{l}{Hydra A} &  0.05384  & $0.88^{\pm{0.02}}$ & $9.5^{\pm{0.2}}$& $17.5^{\pm{0.4}}$ & $8^{+6}_{-4}$\hspace{0.1cm}\tablenotemark{b} & $900^{+}$ & 2.35\\
\multicolumn{2}{l}{Zw3146} & 0.2906  & $8.47^{\pm{0.4}}$ & $91^{\pm{4}}$ & $168^{\pm{8}}$ & $525^{+90}_{-90}$ & 11 & 3.3\\
\multicolumn{2}{l}{Abell 1795} &  0.06326  & $0.85^{\pm{0.02}}$ & $9.1^{\pm{0.2}}$ & $16.7^{\pm{0.4}}$ & $30^{+5}_{-5}$\hspace{0.1cm}\tablenotemark{b} & 100 & 3.3 \\
\multicolumn{2}{l}{Abell 1835} &  0.251982  & $11.41^{\pm{0.4}}$ & $123^{\pm{5}}$ & $226^{\pm{9}}$ & $34^{+43}_{-34}$\hspace{0.1cm}\tablenotemark{b} & $900^{+}$ & 2.35\\
\multicolumn{2}{l}{MS 1455.0+2232} &  0.2578  & $3.36^{\pm{0.4}}$ & $36^{\pm{4}}$ & $66^{\pm{8}}$ & $200^{+70}_{-70}$ & 150 & 3.3 \\
\multicolumn{2}{l}{Abell 2029} & 0.07795  & $0.18^{\pm{0.05}}$ & $1.7^{\pm{0.5}}$ & $3.2^{\pm{0.9}}$ & $30^{+10}_{-13}$\hspace{0.1cm}\tablenotemark{b} & 7 & 3.3\\
\multicolumn{2}{l}{Abell 2052} &  0.03446 & $0.19^{\pm{0.007}}$ & $0.19^{\pm{0.08}}$ & $0.4^{\pm{0.1}}$ & $100^{+20}_{-20}$\hspace{0.1cm}\tablenotemark{c} & \nodata & \nodata \\
\multicolumn{2}{l}{MKW3s} &  0.04531  & $0.01^{\pm{0.01}}$ & $0.1^{\pm{0.1}}$ & $0.2^{\pm{0.2}}$ & $45^{+10}_{-10}$\hspace{0.1cm}\tablenotemark{c}  & \nodata & \nodata\\
\enddata 
\tablenotetext{a}{Assuming constant star formation over 900 Myr.  The subscript indicates the IMF powerlaw slope of the model used.}
\tablenotetext{b}{Voigt \& Fabian (2004).} 
\tablenotetext{c}{Peterson et al. (2003).}
\end{deluxetable}


\end{document}